\documentclass[lettersize,journal]{IEEEtran}
\usepackage{amsmath,amsfonts}
\usepackage{algorithm}
\usepackage{amssymb}
\usepackage{array}
\usepackage[caption=false,font=normalsize,labelfont=sf,textfont=sf]{subfig}
\usepackage{textcomp}
\usepackage{stfloats}
\usepackage{url}
\usepackage{verbatim}
\usepackage{graphicx}
\usepackage{cite}
\usepackage{booktabs} 
\usepackage{algpseudocode}
\usepackage{color}
\usepackage{tikz}
\usepackage{xcolor}
\usepackage{indentfirst,flushend}
\usetikzlibrary{positioning, arrows.meta}

\begin{document}

\title{An Agile Adaptation Method for Multi-mode Vehicle Communication Networks}


\author{Shiwen~He\textit{, Member, IEEE},~Kanghong Chen,~Shiyue Huang, Wei Huang\textit{, Member, IEEE},~and Zhenyu An\textit{, Member, IEEE} 
	
	\thanks{S. He, K. Chen, and S. Huang are with the School of Computer Science and Engineering, Central South University, Changsha 410083, China. S. He is also with the Purple Mountain Laboratories, Nanjing 211111, China (e-mail: shiwen.he.hn@csu.edu.cn).}
	\thanks{W. Huang is School of Computer Science and Information Engineering, Hefei University of Technology, Hefei 230601, China, and also with the Intelligent Interconnected Systems Laboratory of Anhui Province, Hefei University of Technology, Hefei 230601, China (e-mail: huangwei@hfut.edu.cn).}
	\thanks{Z. An is with the Purple Mountain Laboratories, Nanjing 210096, China (email: anzhenyu@pmlabs.com.cn).}
}

\maketitle
\begin{abstract}
This paper focuses on discovering the impact of communication mode allocation on communication efficiency in the vehicle communication networks. To be specific, Markov decision process and reinforcement learning are applied to establish an agile adaptation mechanism for multi-mode communication devices according to the driving scenarios and business requirements. Then, Q-learning is used to train the agile adaptation reinforcement learning model and output the trained model. By learning the best actions to take in different states to maximize the cumulative reward, and avoiding the problem of poor adaptation effect caused by inaccurate delay measurement in unstable communication scenarios. The experiments show that the proposed scheme can quickly adapt to dynamic vehicle networking environment, while achieving high concurrency and communication efficiency.
\end{abstract}

\section{Introduction}
\IEEEPARstart{W}{ith} the continuous evolution of technological development and industrial transformation, the automotive industry is accelerating its development towards intelligence, networking, and high-end \cite{ref1}. Cooperative Connected Automated Driving (CCAD) achieves collaborative sensing, decision-making, and control of intelligent networked vehicles through information exchange among vehicles, roadside units (RSUs), and the cloud, which enables them to advance from single vehicle intelligence to group intelligence \cite{ref2}.

The real-time requirement of information exchanges in CCAD poses a significant challenge to wireless communication systems \cite{ref3}. High-rate, reliable, bidirectional and integrated intelligent network technology with multiple communication modes is important for the realization of intelligent vehicle-road cooperative system, and provides corresponding communication guarantee according to different levels of requirement. In the context of vehicle communication networks (VCNs), the commonly used communication mode is mainly the device-to-device communication (proximity communication, PC5) which coexists with the cellular and Wi-Fi communication systems\cite{ref4}. This implies that vehicles can directly choose the RSUs or cellular networks to exchange the information. In fact, the same message in the VCNs may need to be simultaneously sent to multiple target nodes, including other vehicles, RSUs, and non-vehicle nodes. Furthermore, the transmission content and the received terminals depend on the communication environment, and the intent of driving \cite{ref5}. This communication method expands the coverage area, prevents node overload by dispersing the communication load, and enhances the reliability of data interaction and the adaptability of driving intentions in different communication environments.

In the current multi-mode communication systems, the devices usually select one communication mode to perform the specific services, which cannot meet the actual application requirements in the case with large number of transmission services \cite{ref6}. To address this limitation, the authors of \cite{ref7} proposed an unsupervised deep unrolling framework based on projection gradient descent for solving constrained optimization problems in wireless networks. The authors of \cite{ref8} introduced a neural network-based communication selection mechanism tailored for VCNs. More recently, multi-path transmission and multi-link transmission based on one communication protocol have attracted extensively attention in the field of academia and industry.

Multi-link transmission is commonly employed to ensure the reliable transmission of data with low-latency requirements \cite{ref9}. This technology uses multiple links of different frequencies that contain multiple base service sets to expand the available bandwidth and further reduce latency \cite{ref10}. Multi-path transmission typically involves the use of multiple multi-hop paths, with the primary objective of balancing resource utilization among network nodes to alleviate congestion and decrease delay \cite{ref11}. However, when vehicles or RSUs encounter a surge in transmission tasks due to sudden changes in road conditions, the vehicles must communicate with multiple devices, making it challenging to further reduce transmission delay through parallel multi-link transmission. Multi-path transmission focuses on ensuring data transmission reliability and congestion control \cite{ref12}. Concurrent multi-user transmissions not only risk node congestion but also require frequent status inquiries from other nodes during the transmission process. Moreover, the flexible scheduling of multiple communication modes, as an enhancement to multi-link and multi-path strategies, can further decrease latency and simplify the process of acquiring congestion information from other nodes, thereby addressing the challenges posed by the need for reliable and low-latency communication in VCNs.

Therefore, in order to solve the above problems, this paper proposes a VCN multi-mode selection scheme for bursting services. Specifically, the goal of this paper is to minimize the delay of each communication channel while strictly adhering to the transmission delay constraints inherent to VCN burst services and develops an effective agile adaptation reinforcement learning model (AARLM). In summary, the key technical contributions of this paper are summarized as follows:
\begin{itemize}
	\item{Drawing upon a multi-tiered set of data, encompassing transmission latency requirements, network topologies, and link conditions, this paper explores the concurrent activation of multiple communication modes to facilitate simultaneously multi-user transmissions, thereby addressing the stringent low-latency demands of large-scale burst tasks.}
	\item{This paper proposes an AARLM scheme for multi-mode communication in VCNs scenarios, combining the dynamic environment of driving scenarios, driving intentions and business needs.}
	\item{This paper verifies the feasibility of multi-mode communication in VCNs scenarios through simulation, and proves that AARLM can effectively meet the real-time requirements of vehicle-to-vehicle communication and vehicle-to-road communication.}
\end{itemize}

The rest of this paper is organized as follows: Section II presents the system model and problem formulation. In Section III, a reinforcement learning (RL)-based multi-mode agile adaptation scheme is proposed. Numerical results are presented in Section IV. Finally, Section V concludes the paper.

\begin{figure}[b]
	\centering
	\vspace{-20pt}
	\small
	\begin{subequations}\label{Multi-mode02}
		\begin{align}
			&\min_{\{a(i,j,c)\}}\max_{j\in\widetilde{\mathcal{V}}}\sum_{c\in\mathcal{C}}\frac{a(i,j,c)(F(i,c)+S(i,j,c)M(i,j))}{B(i,j,c)},\label{Multi-mode02a}\\
			\text{s.t.}~~&\sum_{c\in\mathcal{C}}a(i,j,c)=1, \forall j\in\widetilde{\mathcal{V}},\label{Multi-mode02b}\\
			&a(i,j,c)\in\{0,1\}, \forall j\in\widetilde{\mathcal{V}}, \forall c\in\mathcal{C},\label{Multi-mode02c}\\
			&\sum_{c\in \mathcal{C}}\frac{a(i,j,c)(F(i,c)+S(i,j,c)M(i,j))}{B(i,j,c)}\leq D(i,j),\forall j\in\tilde{\mathcal{V}}.\label{Multi-mode02d}
		\end{align}
	\end{subequations}
	\vspace{-10pt}
\end{figure}

\section{System Model and Problem Formulation}
\begin{figure}[htbp]
	\centering
	\includegraphics[width=\linewidth]{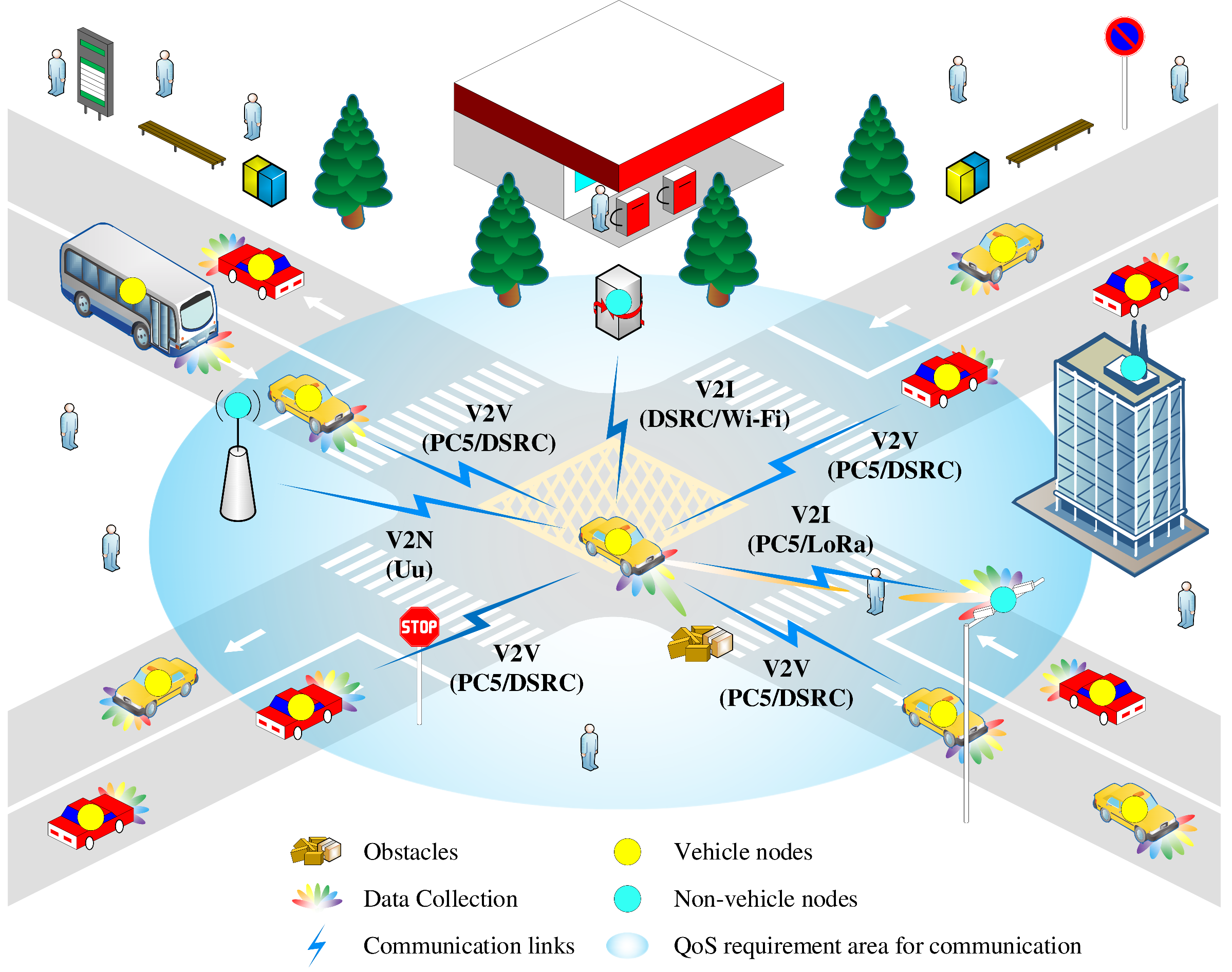}
	\caption{Multi-mode vehicle communication network.}
	\label{Systemmodel}
\end{figure}

This paper considers a multi-mode VCNs, where the vehicles and RSUs have the ability to simultaneously support multiple communication modes (protocols), which usually provide at least two communication mode in the (C-V2X) \cite{ref4}, i.e,
\begin{itemize}
	\item{A short distance direct communication mode proximity communication between vehicles, pedestrians and RSUs.}
	\item{Communication mode between terminals and base stations, which can achieve reliable communication over long distances and over a larger range.}
\end{itemize}
In addition, the dedicated short range communications, Wi-Fi, long range radio are also commonly used communication protocols for vehicles, as illustrated in Fig. \textcolor{red}{\ref{Systemmodel}}.

For the sake of convenience in description, let $\mathcal{V}=\{1,2,\cdots,N\}$ be the set of terminals in the multi-mode VCNs, where $\mathcal{V}$ denotes a collection of multi-mode communication vehicle nodes, RSUs and base stations. Moreover this paper assumes that the multi-mode vehicle node $i$ sends the information-bearing symbols to the other terminals denoted as $\tilde{\mathcal{V}} = \mathcal{V}\backslash\{i\}$.

This paper focuses on the decision-making policy of multi-mode node $i$, aiming to minimize the total transmission latency, as shown in problem \textcolor{red}{\eqref{Multi-mode02}}. Specifically, in formula \textcolor{red}{\eqref{Multi-mode02}}, $a(i,j,c)$ denotes the decision variable of multi-mode node $i$ for multi-mode node $j$, $j\in\widetilde{\mathcal{V}}$ and $c\in\mathcal{C}$. Here, $a(i,j,c)=1$ indicates that the communication protocol $c$ is selected for transmitting data from node $i$ to node $j$, $j\in\widetilde{\mathcal{V}}$, whereas $a(i,j,c)=0$ indicates otherwise. $B(i,j,c)$ is the bandwidth allocated by multi-mode node $i$ for communication with multi-mode node $j$ in mode $c$. $M(i,j)$ and $D(i,j)$ represent the data size that multi-mode node $i$ transmits to multi-mode node $j$ and the maximum tolerable delay between node $i$ and node $j$  respectively. Among them, the data size $M(i,j)$ can be unequal between node $i$ and different nodes. $F(i,c)$ indicates the buffer data at node $i$ for different communication modes $c$. Let \(S(i,j,c)\) be the indicator of whether communication protocol $c$ is supported between node $i$ and node $j$, $i\in\mathcal{V}$, $j\in\widetilde{\mathcal{V}}$, $c\in\mathcal{C}=\{1,2,\cdots,C\}$ with $C$ being the number of communication protocols supporting by the multi-mode nodes in the multi-mode VCNs, i.e.,
\begin{equation}
	\label{Multi-mode01}
	S(i,j,c) \in \{0,1\}, \forall i\in\mathcal{V}, j \in \widetilde{\mathcal{V}},\forall c \in \mathcal{C}.
\end{equation}
In equation \textcolor{red}{\eqref{Multi-mode01}}, $S(i,j,c)=1$ signifies that communication protocol $c$ is supported between multi-mode node $i$ and node $j$; conversely, $S(i,j,c)=0$ indicates the absence of support for protocol $c$ between these two nodes. Unless otherwise specified, every node within the multi-mode VCN is assumed to be a multi-mode node, meaning that there are more than one communication protocols available ($C>1$) ~\footnote{This work assumes that various communication protocol links between communication terminals have been established before using selecting the corresponding mechanisms.}. Constraint \textcolor{red}{\eqref{Multi-mode02a}} means that only one communication mode can be selected between multi-mode node $i$ and multi-mode node $j$. Constraint \textcolor{red}{\eqref{Multi-mode02d}} indicates that the communication delay between multi-mode node $i$ and node $j$ must be less than the maximum tolerable delay.

It is not difficult to see that problem \textcolor{red}{\eqref{Multi-mode03}} is an integer linear programming problem, which is generally challenging to solve directly. As the problem scale increases, particularly with the presence of numerous multi-mode nodes, the number of variables and constraints in the optimization formulation escalates significantly, resulting in a substantial increase in solution time. Therefore, this work proposes the incorporation of RL to solve the agile adaptation problems in different scenarios.

\section{Reinforcement Learning Model for Agile Adaptation Problems}
To achieve low latency, this paper adopts Q-learning \cite{ref13} to train and optimize the vehicle communication decision model. In this section, the agile adaptation scheme is obtained by using the Q-learning method. It is vital to first establish the three fundamental elements of Q-learning: action space, state space and reward function. Subsequently, a tailored model network is designed to cater to the specific requirements of the considered agile adaptation problem. In order to facilitate the design of the model, problem  \textcolor{red}{\eqref{Multi-mode02}} is converted to problem  \textcolor{red}{\eqref{Multi-mode03}} to obtain the minimum delay $\tau$, and constraint \textcolor{red}{\eqref{Multi-mode03b}} indicates that $\tau$ must be greater than the transmission delay between multi-mode node $i$ and any multi-mode node $j$ in $\tilde{\mathcal{V}}$.

\begin{figure}[h]
	\centering
	\vspace{-15pt}
	\small
	\begin{subequations}\label{Multi-mode03}
		\begin{align}
			&\min_{\{a(i,j,c)\}}\tau,\label{Multi-mode03a}\\
			\text{s.t.}~~ &\sum_{c\in\mathcal{C}}\frac{a(i,j,c)(F(i,c)+S(i,j,c)M(i,j))}{B(i,j,c)}\leq \tau,\forall j\in\widetilde{\mathcal{V}},\label{Multi-mode03b}\\
			&\sum_{c\in\mathcal{C}}a(i,j,c)=1, \forall j\in\widetilde{\mathcal{V}},\label{Multi-mode03c}\\
			&a(i,j,c)\in\{0,1\}, \forall j\in\widetilde{\mathcal{V}}, \forall c\in\mathcal{C},\label{Multi-mode03d}\\
			&\sum_{c\in \mathcal{C}}\frac{a(i,j,c)(F(i,c)+S(i,j,c)M(i,j))}{B(i,j,c)}\leq D(i,j),\forall j\in\tilde{\mathcal{V}}.\label{Multi-mode03e}
		\end{align}
	\end{subequations}
	\vspace{-20pt}
\end{figure}

\subsection{Basic Elements of Reinforcement Learning Model}
In the agile adaptation problem \textcolor{red}{\eqref{Multi-mode03}}, the communication network, comprised of multiple multi-mode nodes, is used as the environment of the AARLM, and the multi-mode nodes serve as agents in the AARLM to interact with the environment. To ensure the model's responsiveness to environmental shifts and output the optimal decision-making action according to the current environment, the action space, state space and reward function of the AARLM are described as follows:

State space $\mathcal{S}_i$ is the set of states $s_k$ that are observed by multi-mode node $i$ in the environment, where $k$ is the index of states. The status $s_k$ includes whether the transmission task between multi-mode nodes is complete and the amount of data in the buffer for each communication mode. The task is considered complete if it does not exceed $D(i,j)$.

Action space $\mathcal{A}_i(s_k)$ is the set of actions that can be taken by multi-mode node $i$ in state $s_k$. The action $a(i,j,c)$ represents not only the decision of the multi-mode node $i$, but also the action. For ease of description, $a$ will be used as shorthand for $a(i,j,c)$.

Reward function $R_{i}(s_k,a)$ quantifies the reward associated with action $a \in \mathcal{A}_i(s_k)$ in state $s_k \in \mathcal{S}_i$. The reward function $R_{i}(s_k,a)$ is formulated as \textcolor{red}{\eqref{Multi-mode06}} for $\forall j\in\tilde{\mathcal{V}}$, $\forall c\in\mathcal{C}$.
\begin{equation}
	\label{Multi-mode06}
	{R_i}({s_k},a) = {T_{max}}({s_k},a) - {T_{\max }}({s_{k + 1}},a)
\end{equation}
where ${T_{max}}({s_k},a)$ and ${T_{\max }}({s_{k + 1}},a)$ are respectively the maximum delay before and after action $a(i,j,c)$ is taken in each mode.
\begin{equation}
	\label{Multi-mode07_01}
	{{T_{max}}({s_k},a)} = \mathop {\max }\limits_{j \in \tilde {\cal V}} \sum\limits_{c \in {\cal C}} {\frac{{a(i,j,c)(F(i,c) + S(i,j,c)M(i,j))}}{{B(i,j,c)}}}
\end{equation}

The cumulative reward $R_{i}$ is calculated as
\begin{equation}
	\label{Multi-mode07}
	R_{i}=\gamma\tilde{R}_i(s_k,a)+R_i(s_k,a)
\end{equation}
where $\gamma$ is the discount factor, $\tilde{R}_i(s_k,a)$ represents the reward that multi-mode node $i$ has already received.

$\pi(a|s_k)$ is a mapping policy from the observed states to the actions that will be taken by an agent in those states. For ease of description, $\pi$ in the following content stands for $\pi(a|s_k)$.
\begin{figure}[t]
	\centering
	\includegraphics[width=0.9\linewidth]{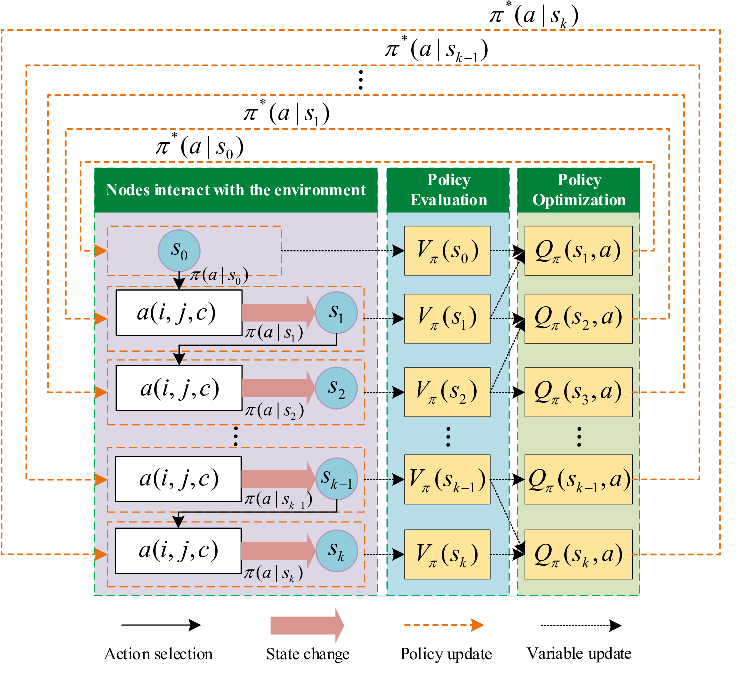}
	\caption{Agile adaptation reinforcement learning model.}
	\label{AARLM}
\end{figure}

\subsection{Framework of proposed Reinforcement Learning Model}
The framework of AARLM illustrated in Fig. \textcolor{red}{\ref{AARLM}} consists of two key steps: policy evaluation and policy optimization. The former module assesses the rewards associated with the chosen actions based on a given policy. The policy optimization module identifies the optimal policy ${\pi ^*}(a\left| s_k \right.)$ by tuning the selection of actions. To evaluate the policy $\pi(a|s_k)$, the multi-mode node $i$ first selects the action $a(i,j,c)$ according to the initial policy $\pi(a|s_k)$ to make the state $s_k$ change, and then calculates the state value. The state value function $V_{\pi}(s_k)$ is defined as
\begin{equation}
	\label{Multi-mode09}
	\begin{aligned}
		{V_\pi }({s_k}) &= \sum\limits_{a \in {{\cal A}_i}({s_k})} \pi  (a|{s_k}) \cdot \\
		&[{R_i}({s_k},a) + \gamma \sum\limits_{{s_{k}}  \in {{\cal S}_i}} {{P}} ({s_{k}} |s_{k - 1},a){V_i}({s_{k}} )]
	\end{aligned}
\end{equation}
where ${{P}({s_k} |{s_{k - 1}},a)}$ represents the probability that multi-mode node $i$ moves to state ${{s_k} }$ after selecting action $a(i,j,c)$ in state $s_k$, and $V_i({{s_k} })$ is one of the state values of next state ${{s_k} }$. In order to evaluate the policy $\pi(a|s_k)$, it is necessary to select actions based on the $\pi(a|s_k)$ and update the state until the termination state is reached.

\begin{algorithm}[t]
	\caption{Policy Iteration of AARLM}
	\label{algorithm01}
	\begin{algorithmic}[1]
		\Require Initializes the state space $\mathcal{S}_i$, state-value function ${V_\pi }({s_0})$, threshold for delay $D(i,j)$, the set of terminals $\mathcal{V}$, the optimal action set ${\cal A}_i^*({c}) = \varnothing$, the number of model iterations $E_1$, the number of task moves $E_2$.
		
		\State // \textbf{Initial assignment of transmission tasks}
		\State Sort tasks in descending order by $M(i,j)$;
		\State Sort modes in descending order by $B(i,j,c)$;
		\State Calculate ${T_{ave}}$ according to (\ref{Multi-mode11});
		\State \textbf{while}
		\State \quad Select the mode $c \in {\cal C}$ with the largest bandwidth, and ${T_c} \le {T_{ave}}$;
		\State \quad \textbf{if} $S(i,j,c)=1$ \textbf{then}
		\State \quad \quad Select the multi-mode node $j$ with the largest $M(i,j)$, then $a(i,j,c)=1$ and $\mathcal{A}_i^*({c})=\mathcal{A}_i^*({c})\cup a(i,j,c)$;
		\State \quad \textbf{end if}
		\State \textbf{until} All the tasks have been arranged;
		\State Update $Q$ table;
		\State // \textbf{Policy Improvement}
		\State \textbf{for episode 1 to $E_1$ do}
		\State \quad Traverse the $Q$ table and select the state $s_k$ with the lowest delay ${T_{max}}({s_k},a)$ as the initial state;
		\State \quad \textbf{for episode 1 to $E_2$ do}
		\State \quad \quad Traverse ${\cal C}$ and select the mode ${c_{tm}}$ with the maximize ${{T_{c}}}$;
		\State \quad \quad Iterate over all tasks in mode ${c_{tm}}$, selecting multi-mode node $j$ with minimizes $\left| {\frac{{M(i,j)}}{{B(i,j,c)}} - \left( {{T_c} - {T_{ave}}} \right)} \right|$;
		\State \quad \quad Traverse ${\cal C}$ and select the mode $c_{rm}$ with the minimizes $T_{c}$ and $S(i,j,c_{rm}) = 1$;
		\State \quad \quad $a(i,j,c_{tm}) = 0$, $a(i,j,c_{rm}) = 1$; 
		\State \quad \quad calculate $R_{i}(s_k,a)$ according to (\ref{Multi-mode06});
		\State \quad \quad \textbf{if} {$R_{i}(s_k,a) > 0$} \textbf{then}
		\State \quad \quad Update $Q$ table and ${\cal A}_i^*({c})$; 
		\State \quad \quad \textbf{end if}
		\State \quad \textbf{end for}
		\State \textbf{end for}
		\State \textbf{return} $min(T_c)$, ${\cal A}_i^*({c})$
	\end{algorithmic}
\end{algorithm}

Through policy evaluation, the value of each state can be determined, thereby providing insight into the quality of each state. However, the existing policy might not be optimal. To further enhance the policy, it's crucial to assess the impact of actions that deviate from the policy in each state. So the function of the state-action value $Q_\pi(s_k,a)$ is designed as follows:
\begin{equation}
	\label{Multi-mode10}
	\begin{aligned}
		Q_\pi(s_k,a)&=R_i(s_k,a)+\\
		&\gamma\sum_{s_k\in\mathcal{S}_i}P(s_k|s_{k-1},a)\sum_{a\in \mathcal{A}_{i}(s_k)}\pi(a|s_k)Q_{\pi}(s_k,a).
	\end{aligned}
\end{equation}
The state-action value $Q$, defined as the long-term reward, represents the expected cumulative discounted reward $\gamma\sum_{s_k\in\mathcal{S}_i}P(s_k|s_{k-1},a)\sum_{a\in \mathcal{A}_{i}(s_k)}\pi(a|s_k)Q_{\pi}(s_k,a)$ for the action $a(i,j,c)$ that is taken by multi-mode node $i$ in the state $s_k$ under policy $\pi$. The objective of the AARLM is to select the optimal policy ${\pi ^*}(a\left| s_k \right.)$ for every state $s_k$ aiming to maximize $Q_\pi(s_k,a)$. The optimal policy for each state is identified through equation \textcolor{red}{\eqref{Multi-mode10-1}}.
\begin{equation}
	\label{Multi-mode10-1}
	{\pi ^*}(a\left| s_k \right.) = \arg \mathop {\max }\limits_{a \in {{\cal A}_i}(s_k)} {Q_\pi }(s_k,a)
\end{equation}

Algorithm \textcolor{red}{\ref{algorithm01}} outlines the procedure for policy update. The initial segment, lines 1 through 11, updates the selected mode based on the initial policy $\pi(a|s_k)$. Specifically, line 6 identifies the mode $c$ subject to the maximum bandwidth condition ${T_c} < {T_{ave}}$, ${T_{ave}}$ is defined as
\begin{equation}
	\label{Multi-mode11}
	{T_{ave}} = {{\sum\limits_{j \in \tilde {\cal V}} {M(i,j)} } \mathord{\left/
			{\vphantom {{\sum\limits_{j \in \tilde {\cal V}} {M(i,j)} } {\sum\limits_{c \in {\cal C},j \in \tilde {\cal V}} {B(i,j,c)} }}} \right.
			\kern-\nulldelimiterspace} {\sum\limits_{c \in {\cal C},j \in \tilde {\cal V}} {B(i,j,c)} }}
\end{equation}

Following that, on line 8, all tasks that support mode $c$ are selected, sorted by data volume from largest to smallest, and transmitted using mode $c$ until ${T_c} > {T_{ave}}$. This iteration process repeats until all tasks have completed mode selection, yielding the initial task assignment plan and the Q-table.

Lines 13 to 25 of Algorithm \textcolor{red}{\ref{algorithm01}} optimize the mode selection policy to ${\pi ^*}(a|{s_k})$. Line 14 starts by locating the state $s_k$ with the lowest delay from the Q-table as the starting state. Subsequently, line 15 identifies the mode $c_{tm}$ with the highest delay in state $s_k$. and selects the task with a size closest to $T_c - T_{ave}$ from those transmitted in mode $c_{tm}$. Line 17 selects mode $c_{rm}$, which is supported by the task and has the smallest delay. The task is then moved to mode $c_{rm}$ in line 19, and the reward $R_{i}(s_k,a)$ is calculated in line 20. If $R_{i}(s_k,a)$ is positive, the Q-table and ${\cal A}_i^*({c})$ are updated until the iteration count reaches $E_1$ and the task move count reaches $E_2$.

\section{Numerical Results}
In this paper, the effectiveness of AARLM is verified by simulation. The configuration of simulation parameters such as the number of multi-mode nodes and the size of transmitted data are shown in Table \textcolor{red}{\ref{table1}}. The effects of the number of multi-mode nodes and the number of modes on the overall delay as well as task completion rate are analyzed respectively, and the existing mode adaptation schemes (random selection) and traditional algorithms (annealing algorithms) are compared.
\begin{table}[htbp]
	\centering 
	\caption{Simulation parameter setting} 
	\label{table1}
	\begin{tabular}{l|ccc} 
		\hline 
		Parameters & Value \\ 
		\hline
		Number of multi-mode nodes & [25,150] \\
		Number of communication modes & [2,5] \\
		Data size (Kbits) &	[5,25] \\
		The maximum tolerable delay (ms) & [1,5] \\
		Bandwidth (MHz) & [10,20,40,80,100] \\
		Delay of initial data in buffer (ms) & [1,10]  \\
		\hline
	\end{tabular}
\end{table}

\begin{figure*}[t]
	\centering
	\begin{minipage}[t]{0.24\linewidth}
		\centering
		\includegraphics[width=\linewidth]{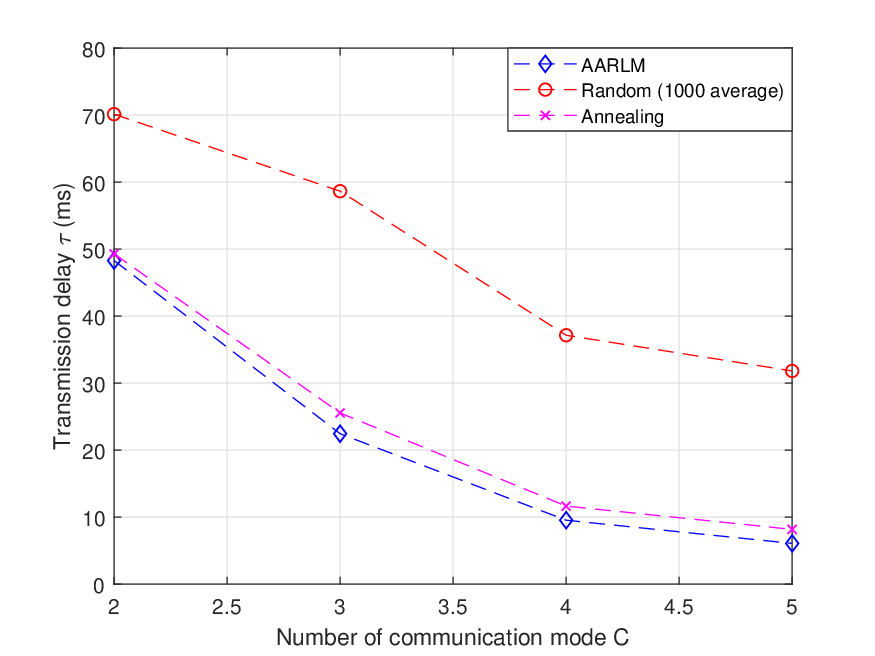}
		\caption{Transmission delay under 100 nodes.}
		\label{fig3}
	\end{minipage}
	\hfill
	\begin{minipage}[t]{0.24\linewidth}
		\centering
		\includegraphics[width=\linewidth]{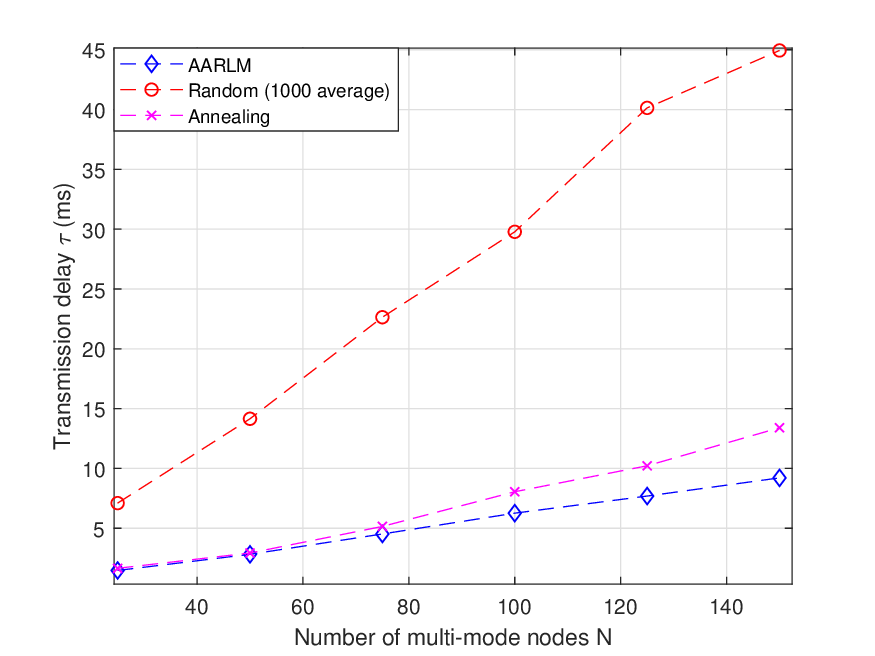}
		\caption{Transmission delay under 5 modes.}
		\label{fig4}
	\end{minipage}
	\hfill
	\begin{minipage}[t]{0.24\linewidth}
		\centering
		\includegraphics[width=\linewidth]{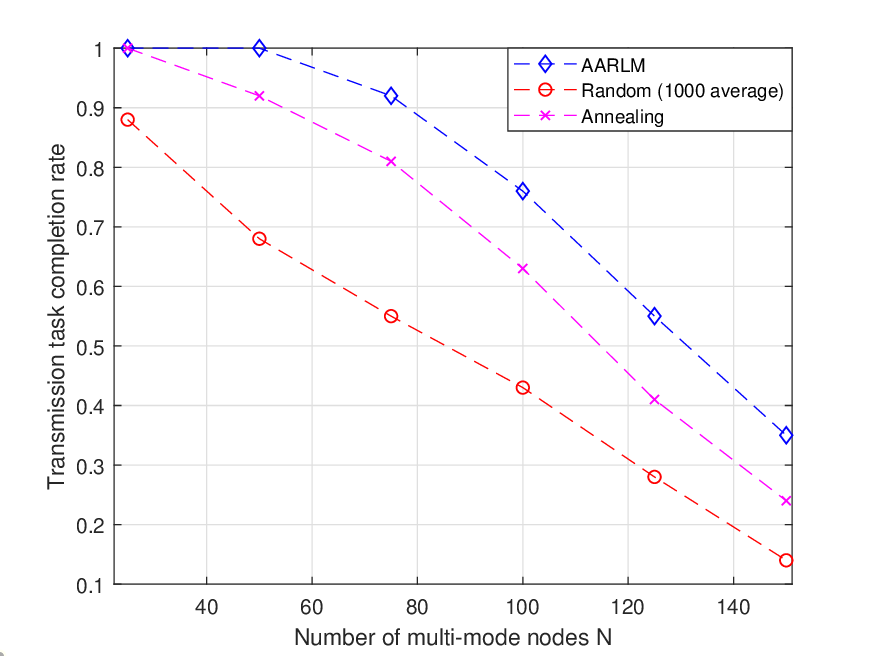}
		\caption{Transmission task completion rate under 5 modes.}
		\label{fig5}
	\end{minipage}
	\hfill
	\begin{minipage}[t]{0.24\linewidth}
		\centering
		\includegraphics[width=\linewidth]{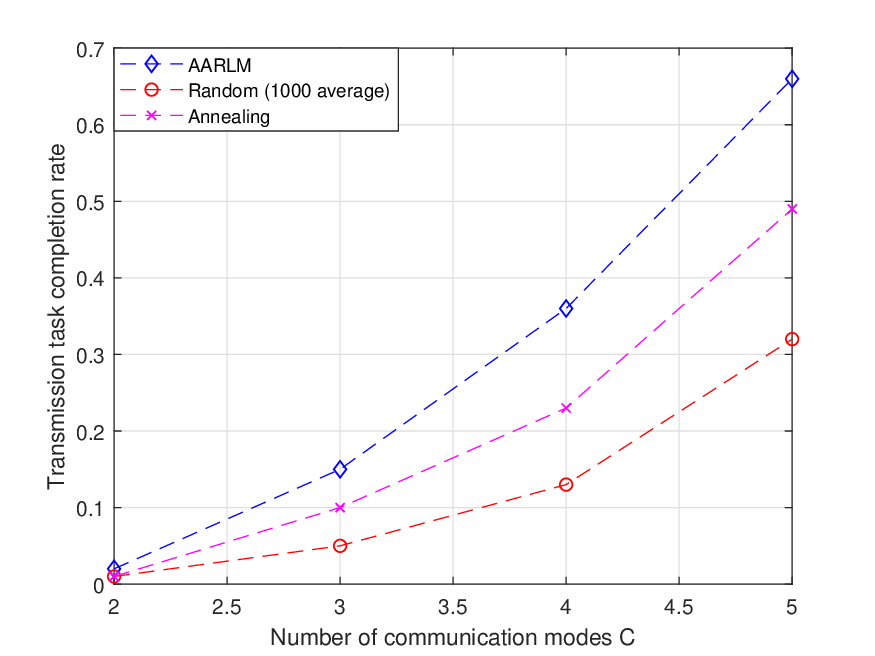}
		\caption{Transmission task completion rate under 100 nodes.}
		\label{fig6}
	\end{minipage}
\end{figure*}

Fig. \textcolor{red}{\ref{fig3}} illustrates the effect of varying communication modes on delay with considering 100 multi-mode nodes. Generally, as the diversity of communication methods broadens, the overall delay of algorithms tends to decrease, with the AARLM achieving the lowest latency. When the bandwidth of the additional modes is 40 MHz or 80 MHz, the reduction in latency is most pronounced initially but then experiences a deceleration. The random algorithm exhibits the highest latency among the evaluated schemes. The delay of AARLM is slightly lower than that of annealing algorithm, and with the increase of communication mode, AARLM has maintained a certain delay advantage over annealing algorithm.

In the case of 5 communication modes, Fig. \textcolor{red}{\ref{fig4}} illustrates the trend in delay changes for various algorithms as the number of multi-mode nodes increases. All algorithms exhibit a general upward trend in delay, and is significantly impacted by small-bandwidth communication modes, leading to the steepest delay increase. Besides, the time delay experiences minor fluctuations due to the inherent randomness in task assignment. As the task volume grows, the complexity of the task allocation scheme intensifies, leading to a widening delay gap between the annealing algorithm and AARLM.

Fig. \textcolor{red}{\ref{fig5}} illustrates the general increase trend in delay changes for various algorithms as the number of multi-mode nodes increases. The random algorithm, in particular, is significantly impacted by small-bandwidth communication modes, leading to the fastest increase in delay. Initially, both the annealing algorithm and AARLM demonstrate satisfactory task completion rates. Nevertheless, when the volume of tasks becomes considerable and exceeds the maximum delay threshold, the task completion rates for both the annealing algorithm and AARLM are observed to decrease sharply. Despite this, AARLM manages to retain a certain degree of superiority.

Fig. \textcolor{red}{\ref{fig6}} shows the variation in task completion rate as the number of communication modes increases, with the task count remains constant at 100. Across all algorithms, the task completion rate rose with more modes, a trend that is accentuated by the higher bandwidth associated with the additional modes, leading to a more rapid enhancement in task completion rates. Notably, AARLM stands out not only in maintaining the highest task completion rate but also in exhibiting the most rapid increase in this crucial metric.

\section{Conclusion}
This paper discussed an RL-based agile adaptation scheme for multi-mode communication in vehicle networks. The preliminary findings have been compelling, revealing that the proposed method has significantly reduced communication delay across all scenarios, outperforming existing schemes. This has proven that the RL-based agile adaptation scheme for multi-mode communication can effectively optimize the communication process and reduce communication delay. Future work will focus on optimizing the framework and incorporating interference factors as well as reliability indicators into multi-mode communication schemes.

\bibliographystyle{IEEEtran}
\bibliography{references}

\begin{thebibliography}{10}
\providecommand{\url}[1]{#1}
\csname url@samestyle\endcsname
\providecommand{\newblock}{\relax}
\providecommand{\bibinfo}[2]{#2}
\providecommand{\BIBentrySTDinterwordspacing}{\spaceskip=0pt\relax}
\providecommand{\BIBentryALTinterwordstretchfactor}{4}
\providecommand{\BIBentryALTinterwordspacing}{\spaceskip=\fontdimen2\font plus
\BIBentryALTinterwordstretchfactor\fontdimen3\font minus
  \fontdimen4\font\relax}
\providecommand{\BIBforeignlanguage}[2]{{%
\expandafter\ifx\csname l@#1\endcsname\relax
\typeout{** WARNING: IEEEtran.bst: No hyphenation pattern has been}%
\typeout{** loaded for the language `#1'. Using the pattern for}%
\typeout{** the default language instead.}%
\else
\language=\csname l@#1\endcsname
\fi
#2}}
\providecommand{\BIBdecl}{\relax}
\BIBdecl

\bibitem{ref1}
K.~Kumar, D.~Zindani, and J.~P. Davim, \emph{{Industry 4.0: Developments
  towards the fourth industrial revolution}}.\hskip 1em plus 0.5em minus
  0.4em\relax Springer, 2019.

\bibitem{ref2}
R.~Xu, H.~Xiang, X.~Han, X.~Xia, Z.~Meng, C.-J. Chen, C.~Correa-Jullian, and
  J.~Ma, ``{The OpenCDA open-source ecosystem for cooperative driving
  automation research},'' \emph{IEEE Trans. Intell. Veh.}, vol.~8, no.~4, pp.
  2698--2711, 2023.

\bibitem{ref3}
B.~H{\"a}fner, V.~Bajpai, J.~Ott, and G.~A. Schmitt, ``{A survey on cooperative
  architectures and maneuvers for connected and automated vehicles},''
  \emph{IEEE Commun. Surv. Tutorials}, vol.~24, no.~1, pp. 380--403, 2021.

\bibitem{ref4}
M.~Ahmed, M.~A. Mirza, S.~Raza, H.~Ahmad, F.~Xu, W.~U. Khan, Q.~Lin, and
  Z.~Han, ``{Vehicular communication network enabled CAV data offloading: A
  review},'' \emph{IEEE Trans. Intell. Transp. Syst.}, vol.~24, no.~8, pp.
  7869--7897, 2023.

\bibitem{ref5}
S.~Zhang, J.~Chen, F.~Lyu, N.~Cheng, W.~Shi, and X.~Shen, ``{Vehicular
  communication networks in the automated driving era},'' \emph{IEEE Commun.
  Mag.}, vol.~56, no.~9, pp. 26--32, 2018.

\bibitem{ref6}
Y.~Fu, C.~Li, F.~R. Yu, T.~H. Luan, and Y.~Zhang, ``{A survey of driving safety
  with sensing, vehicular communications, and artificial intelligence-based
  collision avoidance},'' \emph{IEEE Trans. Intell. Transp. Syst.}, vol.~23,
  no.~7, pp. 6142--6163, 2022.

\bibitem{ref7}
S.~He, S.~Xiong, Z.~An, W.~Zhang, Y.~Huang, and Y.~Zhang, ``{An unsupervised
  deep unrolling framework for constrained optimization problems in wireless
  networks},'' \emph{IEEE Trans. Wireless Commun.}, vol.~21, no.~10, pp.
  8552--8564, 2022.

\bibitem{ref8}
W.~ShangGuan, B.~Shi, B.-G. Cai, J.~Wang, and Y.~Zang, ``{Multiple V2V
  communication mode competition method in cooperative vehicle infrastructure
  system},'' in \emph{2016 IEEE 19th Int. Conf. Intell. Transp. Syst. (ITSC)},
  2016, pp. 1200--1205.

\bibitem{ref9}
T.~Zhou, Y.~Yang, L.~Liu, C.~Tao, and Y.~Liang, ``{A dynamic 3-D wideband GBSM
  for cooperative massive MIMO channels in intelligent high-speed railway
  communication systems},'' \emph{IEEE Trans. Wireless Commun.}, vol.~20,
  no.~4, pp. 2237--2250, 2021.

\bibitem{ref10}
H.~Park and C.~You, ``{Latency impact for massive real-time applications on
  multi link operation},'' in \emph{2021 IEEE Reg. 10 Symp. (TENSYMP)}, 2021,
  pp. 1--5.

\bibitem{ref11}
W.~Liang, S.~Xu, Y.~Liang, Z.~Zhang, and X.~Liu, ``{Researches advanced in
  reliable low latency transmission based on multi-links},'' in \emph{2023 IEEE
  3rd Int. Conf. Electron. Technol., Commun. Inf. (ICETCI)}, 2023, pp. 67--70.

\bibitem{ref12}
K.~Lavanya, R.~Indira, A.~K. Velmurugan, and M.~Janani, ``{Mobility-based
  optimized multipath routing protocol on optimal link state routing in
  MANET},'' in \emph{2023 Int. Conf. Appl. Intell. Sustainable Comput.
  (ICAISC)}, 2023, pp. 1--6.

\bibitem{ref13}
J.~Clifton and E.~Laber, ``{Q-learning: Theory and applications},'' \emph{Ann.
  Rev. Stat. Appl.}, vol.~7, pp. 279--301, 2020.

\end{thebibliography}

\end{document}